\documentclass{article}

\usepackage{arxiv}
\usepackage{lmodern}

\usepackage[utf8]{inputenc} 
\usepackage[T1]{fontenc}    
\usepackage{hyperref}       
\usepackage{url}            
\usepackage{booktabs} 
\usepackage{multirow} 
\usepackage{siunitx} 
\usepackage{amsfonts}       
\usepackage{nicefrac}       
\usepackage{microtype}      
\usepackage{lipsum}
\usepackage{graphicx}
\usepackage{amsmath}
\usepackage{authblk}

\usepackage{booktabs} 
\usepackage{multirow} 
\usepackage{siunitx}
\usepackage{float}
\usepackage{algorithm}
\usepackage{algorithmic}
\usepackage{tabularx} 

\title{Diff-Lung: Diffusion-Based Texture Synthesis for Enhanced Pathological Tissue Segmentation in Lung CT Scans}
\author[1,3]{Rezkellah Noureddine Khiati}
\author[2]{Pierre-Yves Brillet}
\author[3]{Radu Ispas} 
\author[1]{Catalin Fetita} 

\affil[1]{SAMOVAR, Telecom Sud-Paris, Institut Polytechnique de Paris, Evry, France}

\affil[2]{Avicenne Hospital, AP-HP, Bobigny, France}
\affil[3]{Keyrus France, Levallois-Perret, France}

\begin{document}
\maketitle
\begin{abstract}
Accurate quantification of the extent of lung pathological patterns (fibrosis, ground-glass opacity, emphysema, consolidation) is prerequisite for diagnosis and follow-up of interstitial lung diseases. However, segmentation is challenging due to the significant class imbalance between healthy and pathological tissues. This paper addresses this issue by leveraging a diffusion model for data augmentation applied during training an AI model. Our approach generates synthetic pathological tissue patches while preserving essential shape characteristics and intricate details specific to each tissue type. This method enhances the segmentation process by increasing the occurence of underrepresented classes in the training data. We demonstrate that our diffusion-based augmentation technique improves segmentation accuracy across all pathological tissue types, particularly for the less common patterns. This advancement contributes to more reliable automated analysis of lung CT scans, potentially improving clinical decision-making and patient outcomes.

\end{abstract}
\keywords{
Pathological lung tissue, Segmentation, Diffusion based generation}
\section{Introduction}
\label{sec:intro}
Interstitial lung diseases (ILDs) present a significant challenge in medical imaging due to the variability and subtlety of pathological patterns involved (fibrosis, emphysema, ground-glass opacities). These patterns are critical for diagnosing and monitoring various ILDs, yet annotated datasets capturing these conditions are often limited, imbalanced, and expensive to obtain. Although substantial work has been done in the field, most studies have focused on segmenting individual pathologies, such as fibrosis  \cite{DING2022107049,kim2024predicted,Gerard_2021} or emphysema \cite{article,article2} in isolation. Comprehensive segmentation that captures a range of pathological textures within a single model remains underexplored \cite{rennotte2020comparison,10485396}. This lack of a holistic approach limits the generalizability of existing models in clinical practice, where multiple pathologies often coexist and interact within the lung tissue.

 To address the limited and imbalanced datasets available for training, data augmentation techniques have gained prominence, particularly in enhancing segmentation models for medical imaging.

Generative adversarial networks (GANs) \cite{goodfellow2014generativeadversarialnetworks} have gained traction in this area, as they can produce realistic, augmented samples for medical imaging. However, GANs present challenges, including sensitivity to hyperparameters, instability during training, and risks of mode collapse \cite{mescheder2018trainingmethodsgansactually}, which can result in less diverse or overly homogeneous outputs. On the other hand, variational autoencoders (VAEs) \cite{kingma2022autoencodingvariationalbayes} offer a more stable alternative, capable of generating conditioned images. Despite this, VAEs struggle to achieve high-resolution outputs with detailed textures, which are essential for accurate representation of lung pathologies.

Diffusion probabilistic models \cite{ho2020denoisingdiffusionprobabilisticmodels} offer a compelling solution to these challenges. Unlike GANs, diffusion models are less sensitive to hyperparameter tuning and are robust against mode collapse, while also achieving high-resolution generation. In this study, we propose a diffusion model guided by lung masks to synthesize various pathological textures across CT scans. This method enhances dataset diversity, addressing the class imbalance by generating realistic, synthetic samples to balance the representation of each pathology. These synthetic images will subsequently be used to train a baseline (UNet) \cite{ronneberger2015unetconvolutionalnetworksbiomedical} segmentation model, aimed at achieving comprehensive segmentation across multiple lung pathologies in a single framework.

Our contributions in this work lead to the development of DiffLung, a diffusion model for generating lung images with diverse pathological textures. To strengthen DiffLung, we introduce a Class-Balanced Mask Ablated Training strategy (CBMAT), which directs focused learning toward underrepresented pathologies, resulting in a balanced, robust dataset for training a baseline segmentation model. We demonstrate the effectiveness of the proposed augmentation strategy versus standard training of the same baseline model that still uses optimized dataset class balancing.

\section{Method}
\label{sec:methods}
\subsection{Diffusion models}

Following recent advances in diffusion probabilistic models \cite{ho2020denoisingdiffusionprobabilisticmodels}, we leverage their powerful capabilities for texture synthesis in medical imaging. Diffusion models define a Markov chain of diffusion steps that gradually convert a lung CT image $x_0 \in {R}^{c \times h \times w}$ into pure Gaussian noise $x_T$ through a fixed noise schedule. The process then learns to reverse this diffusion to generate new samples.

The forward diffusion process follows a variance schedule $\beta_1,\ldots,\beta_T$ ($0 < \beta_t < 1$), which defines the noise level at each step. At each time step $t$, the process is defined as:

\begin{equation}
    q(x_t|x_{t-1}) = \mathcal{N}(x_t; \sqrt{1-\beta_t}x_{t-1}, \beta_tI)
\end{equation}

This formulation means that at each step, we maintain a fraction $\sqrt{1-\beta_t}$ of the previous image while adding Gaussian noise with variance $\beta_t$. Due to the Markovian nature of this process and the properties of Gaussian distributions, we can express this process in closed form for any time step $t$:

\begin{equation}
    q(x_t|x_0) = \mathcal{N}(x_t; \sqrt{\bar{\alpha}_t}x_0, (1-\bar{\alpha}_t)I)
\end{equation}

where $\bar{\alpha}_t = \prod_{i=1}^t(1-\beta_i)$ represents the cumulative product of noise scheduling coefficients. 

The key innovation of diffusion models lies in their reverse process. While the forward process is fixed and deterministic given the schedule, the reverse process must be learned. This reverse process aims to gradually denoise the image by estimating $p(x_{t-1}|x_t)$. For this, a neural network $\epsilon_\theta$ is used to predict the noise component added during the forward process:

\begin{equation}
    p_\theta(x_{t-1}|x_t) = \mathcal{N}(x_{t-1}; \mu_\theta(x_t,t), \sigma_t^2I)
\end{equation}

The mean of this distribution is computed as:
\begin{equation}
    \mu_\theta(x_t,t) = \frac{1}{\sqrt{1-\beta_t}}\left(x_t - \frac{\beta_t}{\sqrt{1-\bar{\alpha}_t}}\epsilon_\theta(x_t,t)\right)
\end{equation}

where $\epsilon_\theta(x_t,t)$ is our noise prediction network that estimates the noise component in $x_t$. The network takes both the noisy image and the time step as input, allowing it to adapt its predictions based on the noise level.

The training objective is derived from variational inference, but can be simplified to a reweighted denoising score matching objective:

\begin{equation}
    \mathcal{L} = {E}_{t\sim[1,T],x_0,\epsilon}[\|\epsilon_\theta(x_t,t) - \epsilon\|^2]
\end{equation}

where $\epsilon \sim \mathcal{N}(0,I)$ is randomly sampled noise, and $x_t$ is obtained through the reparameterization trick:

\begin{equation}
    x_t = \sqrt{\bar{\alpha}_t}x_0 + \sqrt{1-\bar{\alpha}_t}\epsilon
\end{equation}

This formulation allows us to train the model by sampling random noise and time steps, and then asking the model to predict the noise component. The training process effectively teaches the model to recognize and remove noise at different scales, implicitly learning the data distribution.

During sampling, we start from pure noise $x_T \sim \mathcal{N}(0,I)$ and iteratively apply the learned reverse process to obtain samples from $p(x_0)$. However, for our medical imaging application, sampling from the unconditional distribution $p(x_0)$ is insufficient. 
Following the mask-guided approach introduced in \cite{konz2024segguideddiffusion}, rather than sampling from the unconditional distribution $p(x_0)$, the diffusion process can be controlled through spatial anatomical guidance. This is achieved by conditioning the generation of a $c$-channel image $x_0 \in {R}^{c\times h\times w}$ on a multi-class anatomical mask $m \in \{0,\ldots,C-1\}^{h\times w}$, where each value represents a specific tissue class among $C$ possible classes (including background). This transforms the generative objective into sampling from the conditional distribution $p(x_0|m)$.

A key insight from this approach is that while the forward noising process $q(x_t|x_{t-1})$ remains unchanged since noise addition is independent of anatomical structures, the reverse process must be modified to account for the mask information. This modification affects both the reverse process, transforming from $p_\theta(x_{t-1}|x_t)$ to $p_\theta(x_{t-1}|x_t,m)$, and the noise prediction network $\epsilon_\theta$. Based on the evidence lower bound (ELBO) derivation for this conditional generation setup, the loss function becomes:

\begin{equation}
    \mathcal{L}_m = {E}_{t\sim[1,T],(x_0,m),\epsilon}[\|\epsilon - \epsilon_\theta(x_t, t|m)\|^2]
\end{equation}

where $\epsilon \sim \mathcal{N}(0,I)$ is randomly sampled noise, and the noise prediction network $\epsilon_\theta$ now takes the mask $m$ as an additional input to guide the denoising process.

\subsection{Class-Balanced Mask Ablated Training (CBMAT)}

In medical image analysis, mask-guided models often face the partial label problem, where missing annotations can mislead the model into treating unlabeled regions as background rather than potential pathologies. To address this challenge, \cite{konz2024segguideddiffusion} proposed a mask-ablated training (MAT) strategy which  randomly removes mask classes with uniform probability ($2^{C-1}$ combinations).
However, this uniform sampling fails to account for the inherent class imbalance in medical datasets. To overcome this, we propose a class-balanced mask ablated training (CBMAT) strategy that takes into account pathology prevalence in the training dataset. Specifically, we compute ablation probabilities inversely proportional to class frequencies:

\begin{equation}
    P(\text{ablation}|\text{class}_i) = 1 - f_i
\end{equation}

where $f_i$ represents the frequency of class $i$ in the training dataset. This formulation ensures that commonly occurring pathologies are ablated more frequently during training, forcing the model to learn robust representations of rarer conditions.

Additionally, to maintain anatomical consistency, we exclude the surrounding lung tissue class from ablation, ensuring the generated images retain correct structural boundaries. 
Furthermore, we incorporate a cosine annealing schedule for the ablation probabilities:

\begin{equation}
    P_t(\text{ablation}) = P_{\text{initial}} \cdot \frac{1 + \cos(\pi t/T)}{2}
\end{equation}

where $t$ is the current training iteration and $T$ is the total number of iterations. This scheduled approach gradually reduces the ablation frequency, allowing the model to first learn robust feature representations before fine-tuning on complete masks.

\begin{algorithm}
\caption{Class Balanced Masked Ablation with Cosine Annealing}
\begin{algorithmic}[1]
\REQUIRE Number of mask classes $C$, dataset $p(x_0, m)$, class frequencies $\{f_c\}_{c=1}^{C-1}$, initial ablation probability $\delta_0$
\ENSURE Model parameters $\theta$

\STATE Initialize $\delta_c \leftarrow \delta_0 \times (1 - f_c)$ for each class $c$ to inversely balance ablation based on frequency

\REPEAT
    \STATE Sample $(x_0, m) \sim p(x_0, m)$
    \FOR{$c = 1, \dots, C-1$}
    
        \IF{$c$ is not the surrounding lung tissue} 
            \STATE Sample $\delta \sim \text{Uniform}(0, 1)$
            \IF{$\delta < \delta_c$}
                \STATE Set $m[m = c] \leftarrow 0$ \COMMENT{Ablate class $c$ if sampled}
            \ENDIF
            \STATE Update $\delta_c \leftarrow \delta_c \times 0.5 \times (1 + \cos(\pi \cdot \frac{\text{epoch}}{\text{max\_epoch}}))$ \COMMENT{Cosine annealing for class ablation probability}
        \ENDIF
    \ENDFOR
    \STATE Sample $\epsilon \sim \mathcal{N}(0, I_n)$; $t \sim \text{Uniform}(\{1, \dots, T\})$
    \STATE $x_t \leftarrow \sqrt{\bar{\alpha}_t} x_0 + \sqrt{1 - \bar{\alpha}_t} \epsilon$
    \STATE Update $\theta$ with $\nabla_\theta \| \epsilon - \epsilon_\theta(x_t, t | m) \|^2$
\UNTIL{converged}

\end{algorithmic}
\end{algorithm}

\subsection{Mask Generation Process}
\label{augementation_section}
The DiffLung model will be employed in data augmentation during training a baseline segmentation model. The objective is to compensate for underrepresented classes in the dataset by generating synthetic CT images with the desired spatial distribution of the disease lung patterns. In this respect, the available annotated lung masks are modified to include the required classes as follows. We compute the frequency of each pathology class within this dataset and then apply an iterative balancing algorithm which generates additional samples for less-represented classes. For each iteration, we identify regions belonging to the least frequent class and apply random augmentations such as rotation, copy-paste to a different location, and dilation. These transforms are applied with cascading probabilities, ensuring controlled variability. In order to avoid replacement of existing diseased regions in the lung mask, we constrain class augmentation only in the normal lung region. The process continues until the dataset reaches a threshold of class balance, enhancing the effectiveness of training for our segmentation model.

\section{Experiments}
\subsection{Dataset}
The dataset for this study was collected at Avicenne Hospital, Bobigny, France and includes 156 patients with a total of 2266 CT slices. Of these, 137 patients (2076 slices) are used for training and validation and 19 patients (190 slices) for testing. The class distribution in the annotated lung masks is the following: 12.3\% normal lung tissue, 0.3\% emphysema, 3.8\% interstitial lung disease (ILD, combining fibrosis and ground glass as a single entity), and 83.6\% non-lung regions. 

\subsection{Generative model}
We employed the segmentation-guided diffusion model \cite{konz2024segguideddiffusion} implementation, modifying the mask-ablated strategy with our Class-Balanced Mask Ablation Technique (CBMAT) to enhance focus on underrepresented classes. For preprocessing, we standardized the CT lung images to the Hounsfield Unit range of [-1000, 1000]. We isolated the surrounding tissue and lung regions using the corresponding masks, setting all other pixels to -1000 to prevent unnecessary reconstruction during denoising. We employed the DDIM scheduler \cite{song2022denoisingdiffusionimplicitmodels} for faster inference with 1000 steps, and trained the model for 400 epochs.
\subsection{Quantitative Results}

We conducted a quantitative evaluation of our model, DiffLung with the CBMAT algorithm, comparing it against other segmentation-guided methods, including the mask-ablated approach and GAN-based architectures such as Pix2Pix \cite{isola2018imagetoimagetranslationconditionaladversarial} and SPADE GAN \cite{park2019semanticimagesynthesisspatiallyadaptive}. Given the importance of capturing fine details in medical imaging, we chose Structural Similarity Index Measure (SSIM) and Peak Signal-to-Noise Ratio (PSNR) as our evaluation metrics \cite{1284395}, as they are well-suited for assessing image quality in terms of structural accuracy and noise level. SSIM evaluates structural similarity between images, while PSNR measures the fidelity of image reconstruction, both critical for medical image generation.
Our model outperformed other methods, as shown in the table below, achieving higher SSIM and PSNR scores, which indicates superior quality in preserving anatomical details and reducing noise. These results demonstrate the effectiveness of DiffLung with CBMAT in generating realistic and diverse pathological textures compared to traditional GAN-based models.
\begin{table}[h!]
\centering
\begin{tabularx}{\textwidth}{l >{\centering\arraybackslash}X >{\centering\arraybackslash}X}
\toprule
& \multicolumn{2}{c}{Quantitative Metrics} \\
\cmidrule(lr){2-3}
\textbf{Models} & \textbf{PSNR $\uparrow$ (dB)} & \textbf{SSIM $\uparrow$} \\
\midrule
Pix2Pix                & 17.56 & 0.61 \\
SPADE                  & 19.20 & 0.66 \\
Seg-Guided-Diffusion   & 19.64 & 0.68 \\
Diff-Lung (CBMAT)      & \textbf{20.70} & \textbf{0.71} \\
\bottomrule
\end{tabularx}
\caption{Quantitative comparison of PSNR and SSIM scores for different models.}
\label{tab:psnr_ssim}
\label{difmodels}
\end{table}

\begin{figure}[h!]

  \centering
  \centerline{\includegraphics[scale=0.6]{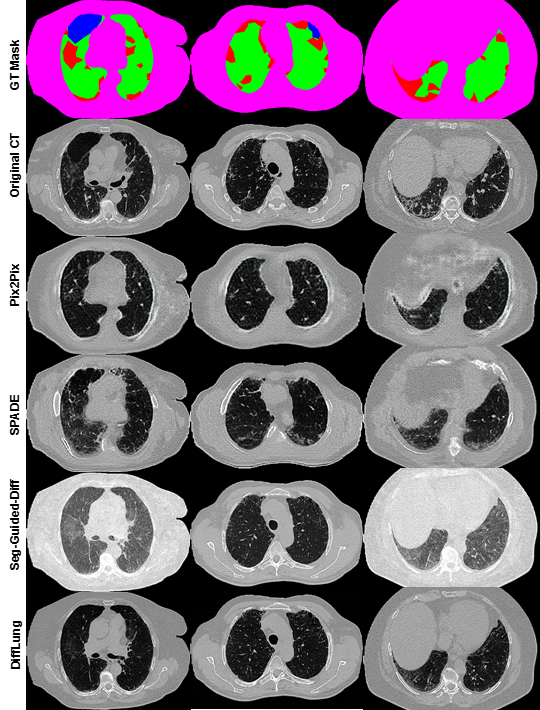}}
  
\caption{Qualitative comparison for pathological texture synthesis in lung CT images (Green: Healthy Tissue, Blue: Emphysema, Red: Fibrosis, Pink: thorax cage and mediastinum).}
\label{fig:res}
\end{figure}
\vspace{-0.5 cm}

\subsection{Segmentation model}

We have investigated the effectiveness of the proposed generative data augmentation approach by comparing the performance of a baseline segmentation model trained with and without synthetically augmented images. Taking into account the generative models results in Table \ref{difmodels}, we also tested the performance gain when using DiffLung versus segmented-guided diffusion \cite{konz2024segguideddiffusion} as generative approach. 
 We chose UNet as baseline architecture trained using CyclicLR scheduler \cite{smith2017cyclicallearningratestraining}  with an exponential range and a Lookahead \cite{zhang2019lookaheadoptimizerksteps} wrapper optimizer, where SGD was used as the base optimizer to avoid local minima that could result from other classes especially affecting the underrepresented emphysema class. 
 We trained three different models: 1) a reference model trained with classic data augmentation strategy - {\it ref-UNet} \cite{rennotte2020comparison}; 2) a model trained with segmented-guided diffusion based data augmentation - {\it Seg-Guided-UNet}; and 3) a model trained with proposed DiffLung based data augmentation - {\it DiffLung-UNet}.
 The generative data augmentation used for the latter two models produced 534 additional slices, resulting in a class representation frequency of 6\% for emphysema and 22\% for fibrosis.
\subsection{Results and discussion}

The quantitative segmentation results of the baseline and generative-augmented models on the test database are synthesized in Table \ref{perf} in terms of Dice coefficients per lung tissue class. They show superior performance of the proposed DiffLung-based augmentation approach, also versus segmented-guided diffusion model which validates the effectiveness of the proposed data augmentation approach.

It is worth noting that the DiffLung model was trained using the same database as for the training the segmentation model. While this choice was imposed by the availability of the annotated data at our hand, we believe that the highly imbalanced lung tissue classes, even partially compensated by the CBMAT strategy, has limited the performance of the generative models (especially for emphysema, as seen in Fig. \ref{fig:res}). Also, having  unifying fibrosis and ground glass in a single ILD class will restrict the distinction of these patterns in the generated images. Future work will consider increasing the training dataset of the DiffLung model by including new samples of less represented classes, mainly emphysema, and separating the ILD class into fibrosis and ground glass.

\begin{table}[h]
\centering
\begin{tabularx}{\textwidth}{l >{\centering\arraybackslash}X >{\centering\arraybackslash}X >{\centering\arraybackslash}X}
\toprule
& \multicolumn{3}{c}{Dice Scores per Class} \\
\cmidrule(lr){2-4}
\textbf{Models} & \textbf{D\textsubscript{Healthy} $\uparrow$} & \textbf{D\textsubscript{Emphy} $\uparrow$} & \textbf{D\textsubscript{Fibrosis} $\uparrow$} \\
\midrule
ref-UNet         & 0.91 & 0.60 & 0.72 \\
Seg-Guided-UNet  & 0.90 & 0.73 & 0.76 \\
DiffLung-UNet    & \textbf{0.92} & \textbf{0.80} & \textbf{0.81} \\
\bottomrule
\end{tabularx}
\vspace{5pt}
\caption{Quantitative comparison of Dice scores per class for different models.}
\label{perf}
\end{table}

\section{Conclusion}

This paper proposed DiffLung, a diffusion model for generating lung images with various diseased patterns exploiting a class-balanced mask ablated strategy to compensate for underrepresented classes in the training dataset. DiffLung demonstrated its ability to increase the performance of a lung texture segmentation model when used as generative data augmentation. 

\section{Acknowledgments}
\label{sec:acknowledgments}
This study was funded by the French National Research Agency (under the ANR MLQ-CT project).
\newpage
\bibliographystyle{unsrt} 
\bibliography{references}

\end{document}